\documentclass[10pt,conference]{IEEEtran}
\IEEEoverridecommandlockouts

\usepackage[T1]{fontenc}
\usepackage{cite,graphicx,xspace,xurl}
\usepackage{microtype}
\usepackage{tikz}
\usetikzlibrary{arrows.meta,positioning}
\usepackage[hidelinks]{hyperref}

\newcommand{\repgen}{\textsc{RepGen}\xspace}
\newcommand{\fname}[1]{\textit{#1}}
\newcommand{\toolURL}{\url{https://github.com/mehilshah/RepGen}}
\newcommand{\packageURL}{\url{https://pypi.org/project/repgen-ai/}}

\newcommand{\videoURL}{\url{https://youtu.be/ZPhPXEDfrRU}}

\begin{document}

\title{RepGen: Reproducing Deep Learning Bugs through\\
Context-Aware Code Generation and Verification}

\author{
\IEEEauthorblockN{Mehil B Shah}
\IEEEauthorblockA{
\textit{Faculty of Computer Science}\\
\textit{Dalhousie University}\\
Halifax, Canada\\
shahmehil@dal.ca}
\and
\IEEEauthorblockN{Mohammad Masudur Rahman}
\IEEEauthorblockA{
\textit{Faculty of Computer Science}\\
\textit{Dalhousie University}\\
Halifax, Canada\\
masud.rahman@dal.ca}
\and
\IEEEauthorblockN{Foutse Khomh}
\IEEEauthorblockA{
\textit{Dept. of Computer \& Software Eng.}\\
\textit{Polytechnique Montr\'eal}\\
Montr\'eal, Canada\\
foutse.khomh@polymtl.ca}
}

\maketitle

\begin{abstract}

Reproducing deep learning (DL) bugs is challenging because failures can involve source code, models, data, dependencies, and execution environments. Incomplete bug reports further complicate reproduction by omitting essential details. We present \repgen, an open-source tool for reproducing DL bugs through context-aware code generation and iterative verification. Given a bug report and repository, \repgen explores the codebase to construct a learning-enhanced context, develops a reproduction plan, and generates candidate scripts. It refines candidates using static checks and LLM feedback to improve their correctness and relevance. When enabled, the tool prepares the execution environment, runs the script, and assesses its output against the reported failure. \repgen provides an interactive CLI supporting local and hosted LLMs, configurable execution budgets, and inspectable artifacts, along with a GitHub Action for automated issue processing. The underlying approach reproduced 85 of 106 real-world DL bugs (80.19\%). A controlled study with 27 developers showed a 23.35\% improvement in reproduction success and a 56.8\% reduction in reproduction time. The demonstration illustrates the workflow using a real-world input-sanitization bug. Tool: \toolURL. Video: \videoURL.
\end{abstract}


\vspace{-1em}
\section{Introduction}
\label{sec:introduction}

Deep learning (DL)-based software systems combine source code with learned models, training data, and execution environments. Their failures can stem not only from the source code but also from the context or environment~\cite{shah2025towards}. To investigate a reported failure, developers often need to reconstruct these execution conditions and reproduce the erroneous behavior. However, bug reports often miss important details, such as required dependencies, configurations, and reproduction steps, making fault reproduction a challenging task~\cite{rahman2020nonreproducible}. A recent study reported successful manual reproduction for only about 3\% of the examined DL bugs, underscoring their inherent challenges~\cite{shah2025towards}.

\looseness=-1
Existing techniques aim to reproduce several types of software failures, but their assumptions do not always hold for DL systems~\cite{shah2026imitation}. For example, the context connected to a deep learning failure can span model architectures, training loops, data processing pipelines, and dependencies. Hence, techniques~\cite{zhao2022recdroid+} that replay GUI events for bug reproduction are not applicable to DL bugs (i.e., no graphical interface). Many DL bugs also manifest as incorrect output or abnormal training behavior without an exception~\cite{tambon2024silent}. Hence, techniques relying only on program crashes or explicit error messages cannot reproduce such DL bugs. Large language models (LLMs) have found numerous applications in software engineering and can be used to generate reproduction scripts against a bug report. However, a generated script (a.k.a., code snippet) may hallucinate about a project-specific API, omit a required model initialization step, or fail because of an incompatible dependency version rather than the reported bug~\cite{shah2026imitation}. Hence, to reproduce a DL bug effectively, we need a candidate that captures the project-specific context and verifies that its behaviour matches the erroneous behaviour described in the report.

In this paper, we present \repgen, a novel agentic solution that helps reproduce deep learning bugs through context-aware code generation and verification using LLM-based agents. Given a bug report and its repository, \repgen navigates the repository to find relevant classes, methods, training loops, and dependencies, and constructs a learning-enhanced context. Then, it leverages the context not only to formulate a reproduction plan but also to generate candidate code. Our tool then iteratively refines the candidate based on four types of verification feedback, including syntax, code quality, semantic relevance, and approximate failure behaviour. Once the candidate code (i.e., reproduction script) is finalized and execution is enabled, \repgen prepares the execution environment by installing the required dependencies, runs the candidate and captures its execution output. Finally, it compares this output with the reported behaviours and generates a detailed reproduction report with diagnostic evidence for developer review. Our tool can be accessed through an interactive command-line interface (CLI) and a GitHub Action.

To summarize, \repgen equips its users with the following capabilities during DL fault reproduction:

\noindent\textbf{(a) Script generation from bug reports:} Users can provide a bug report and its repository to obtain a reproduction script without manually locating relevant code or writing the script from scratch.

\noindent\textbf{(b) Environment setup and script execution:} Users can ask \repgen to install required dependencies, run the generated script, and assess whether its output matches the reported failure.

\noindent\textbf{(c) Flexible configuration and access:} Users can select local or hosted models, set exploration and retry limits, and run the tool through an interactive CLI or a GitHub Action.

\noindent\textbf{(d) Inspectable results:} Users can review saved scripts, verification reports, and execution logs to understand reproduction outcomes, repeat executions, and investigate unsuccessful attempts.

\section{Working Methodology}
\label{sec:methodology}

\begin{figure*}[t]
\centering
\includegraphics[width=0.65\textwidth]{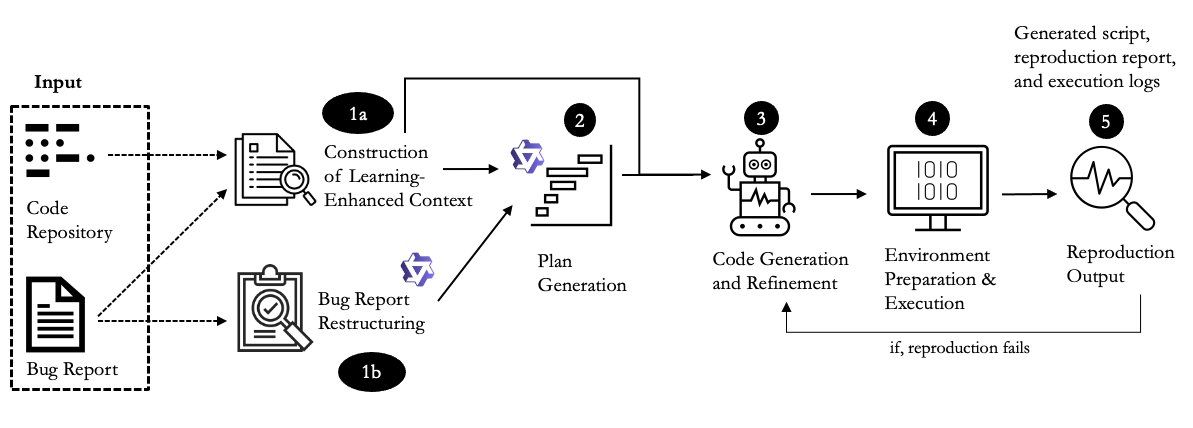}
\vspace{-0.5em}
\caption{Overview of \repgen's underlying approach}
\vspace{-1.5em}
\label{fig:pipeline}
\end{figure*}

\looseness=-1
\repgen combines repository exploration, reproduction planning, code generation, and verification within one workflow. Figure~\ref{fig:pipeline} summarizes the underlying methodology of our developed solution. We describe the major steps of our approach below.

\noindent\textbf{(a) Report Restructuring and Context Construction:} \repgen accepts a GitHub issue or a local bug report and the corresponding repository as inputs. Then it identifies the affected version of the codebase based on the provided commit or the last commit before the bug report submission. It also organises the report into symptoms, expected behaviour, and reproduction steps (Figure~\ref{fig:pipeline}, Step~1b).

\begin{figure*}[t]
\centering
\includegraphics[width=0.6\textwidth]{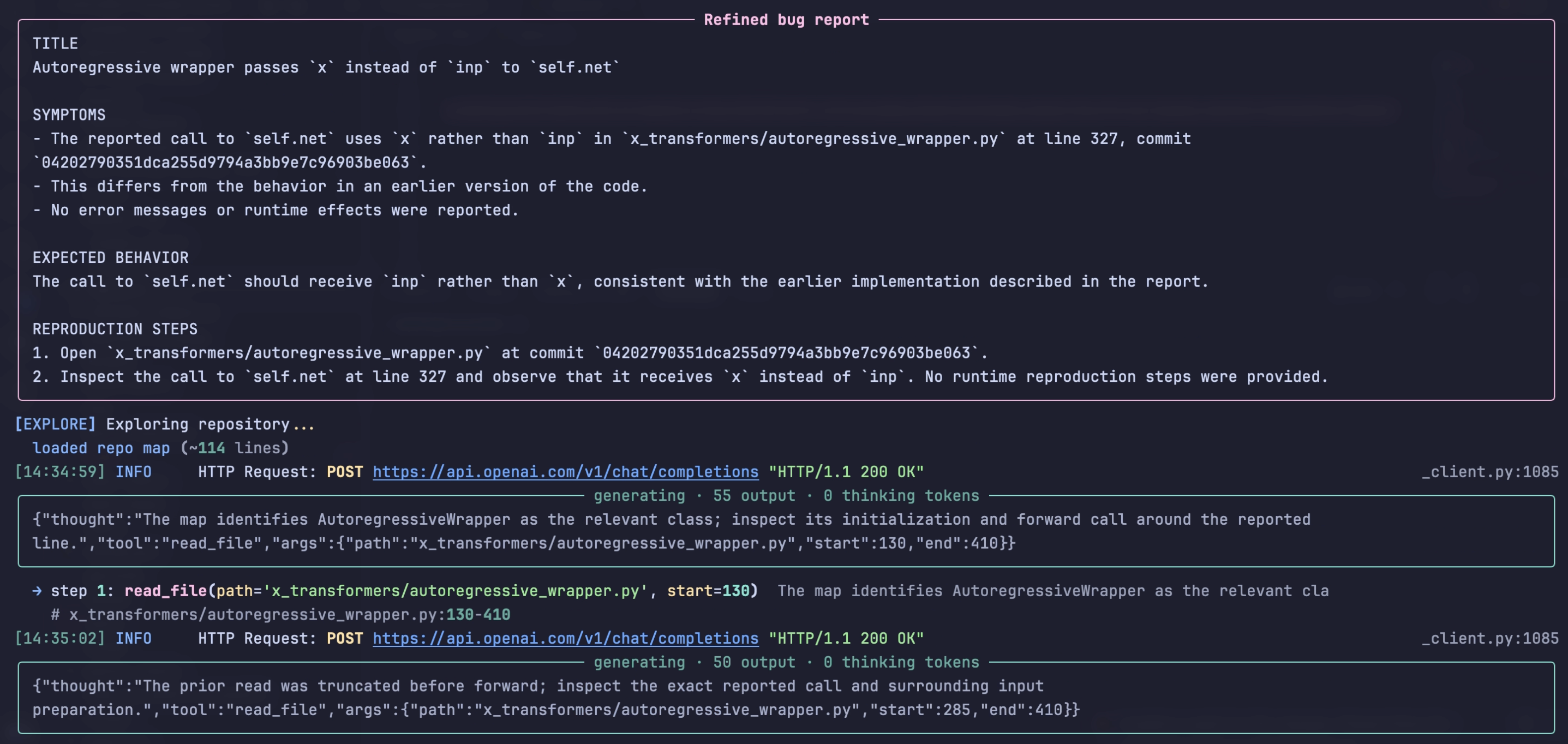}
\par\smallskip
\vspace{-0.5em}
{\footnotesize (a) Report preparation and repository exploration.}
\par\medskip
\includegraphics[width=0.6\textwidth]{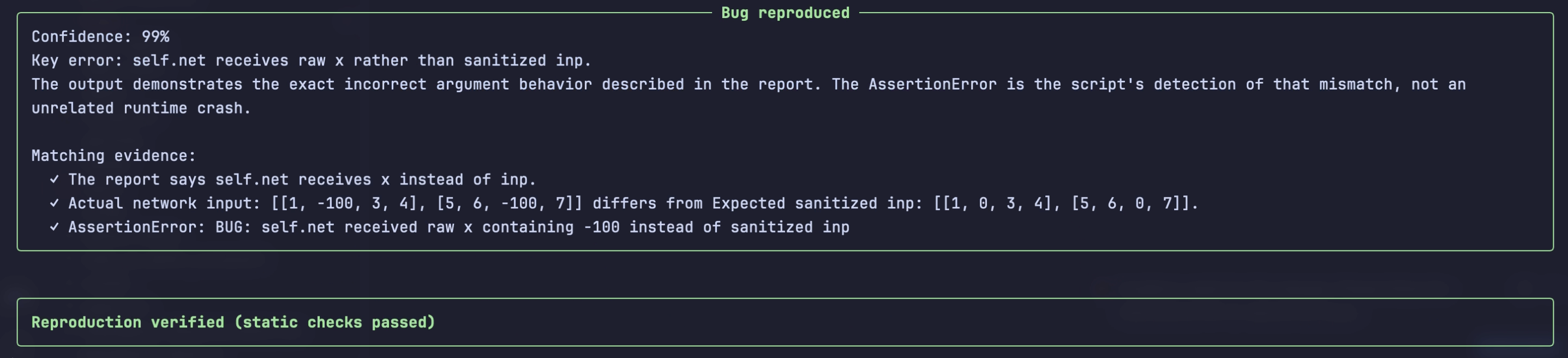}
\par\smallskip
\vspace{-0.5em}
{\footnotesize (b) Reproduction assessment and supporting evidence.}

\caption{\repgen reproducing an input-sanitization bug.}
\label{fig:method-screenshots}
\vspace{-1em}
\end{figure*}

To recover missing context in the bug report, \repgen constructs a map of the code repository using abstract syntax trees (ASTs) to identify modules, classes, methods, and signatures. Then our tool leverages the map and employs a BM25-based search engine and a text-ReAct agent~\cite{yao2023react} to systematically explore the codebase and identify the code relevant to the target fault. After each action, the agent follows a think-action-observe cycle and captures relevant information, as shown in Figure~\ref{fig:method-screenshots}(a). The restructured bug report identifies an incorrect argument (passed to \textit{self.net}) and missing reproduction steps. The agent identifies \textit{AutoregressiveWrapper} as the argument type, narrows the search, and supplements the existing information. This exploration produces a learning-enhanced context encompassing definitions of the relevant classes and methods, training code, and dependencies relevant to the reported failure (Figure~\ref{fig:pipeline}, Step~1a).

\noindent\textbf{(b) Candidate Generation and Verification:} Using the restructured report and the constructed context, \repgen develops a reproduction plan outlining the steps towards failure (Figure~\ref{fig:pipeline}, Step~2). The reproduction agent then generates a candidate reproduction script based on the report, context, and plan (Step~3). It also extracts a failure signature, comprising the expected exception, corresponding error message, and associated APIs when available.

Then, \repgen iteratively refines the candidate based on four types of verification feedback (Step~3). Syntax parsing checks whether the script follows Python grammar, and bytecode compilation checks whether it can be compiled. Focused Pylint checks identify static errors such as undefined names or deprecated APIs. A pseudo-symbolic check also compares the candidate's AST with the constructed repository map (Step~1) for missing imports, required arguments, unsupported keywords, and other suspicious statements.

LLM feedback complements the above checks by estimating the candidate's relevance to the reported fault (Step~3) and by approximating the candidate's output and failure symptoms against the report (Step~3)~\cite{shah2026imitation}. After passing the verification checks or reaching the maximum number of refinement attempts, the candidate proceeds to the next step.

\looseness=-1
\noindent\textbf{(c) Environment Preparation and Execution:} \repgen allows the user to execute the candidate code generated above. When the execution option is enabled, our tool identifies the candidate's third-party dependencies through its import statements and the repository's requirements. It also inspects available computing resources to plan the execution. The environment manager creates a Python virtual environment and installs the project's dependencies. Then it executes the code with a timeout based on the execution plan. It captures standard output, standard error, exit status, and the execution traceback (Figure~\ref{fig:pipeline}, Step~4).

\noindent\textbf{(d) Reproduction Assessment and Reporting:} \repgen compares the execution outcome with the reported failure signature captured above (Step~3) by employing an LLM. It involves matching the exception types, message text, and API names. The LLM also compares the reported erroneous behaviour in the bug report, including symptoms, with the execution outcome. The assessment returns either \emph{reproduced}, when the evidence supports reproducing the reported failure, or \emph{not reproduced}, when the execution does not match. Each verdict also includes a confidence score and supporting evidence. Figure~\ref{fig:method-screenshots}(b) shows a successful reproduction: the assertion failure detects the reported input-sanitization bug. \repgen saves each run, including the bug report, generated script, verification reports, and available execution logs, in a separate directory and records the verdict with supporting evidence (Figure~\ref{fig:pipeline}, Step~5).

\section{User Interfaces and Working Modes}
\label{sec:interfaces}

\repgen supports two working modes: interactive and non-interactive, and provides two user interfaces: CLI and GitHub Action, as follows.

\noindent\textbf{(a) Interactive Command-Line Interface:} Users can start the CLI workflow by executing the command -- \textit{repgen run}. The interactive interface requests missing inputs and guides model selection through useful commands such as \textit{repgen backends} and \textit{repgen models}. Users can also set the bug-inducing commit, exploration budget, generation attempts, and execution retries to match available resources (e.g., machine specifications, token budget). During a run, the CLI displays exploration actions, verification findings, resource decisions, and execution results. Model responses can stream as they arrive, while \textit{--verbose} displays detailed installation and program output. The \textit{--no-execute} option stops the workflow after generation and verification, allowing users to inspect the candidate before running it. Thus, the CLI is suitable for interactive, individual investigations and budget-aware execution.

\noindent\textbf{(b) GitHub Integration:} \repgen can be integrated into GitHub Actions to support automated triage of DL bug reports. A user can add \repgen to the GitHub Actions workflow as a job by configuring the backend credential as a repository secret and creating the \textit{needs-repro} label. Applying this label invokes \repgen with the issue and repository as inputs. Once the action is complete, the workflow returns \repgen's verification report as an issue comment and uploads the result directory as an artifact. The user can inspect the generated script and findings alongside the report. Runtime execution is disabled by default but can be requested through the \textit{execute} input. Given the challenge of missing information in bug reports, this integration provides complementary items for subsequent debugging steps.

\noindent\textbf{Availability and Installation:} \repgen is available on PyPI at \packageURL. Users can install the packaged tool with \textit{pip install repgen-ai}, without cloning or building its source. The source code and usage instructions are available in the following repository -- \toolURL. The repository also includes a Dockerfile that lets users build a container image of the tool.

\vspace{-0.1em}
\section{A Use Case Scenario}
\label{sec:use-case}

Let us consider Alice, an AI practitioner who encounters issue~\#331 in the \textit{x-transformers} repository. Invalid values reach a neural network despite the presence of an input-sanitization step. Specifically, the affected program passes a tensor containing the sentinel value \textit{-100} to the neural network (i.e., \textit{self.net}) without raising an exception or crashing. Thus, Alice must exercise the affected code path and inspect the input passed to the model to reproduce the reported failure.

To investigate the bug, Alice installs \repgen, selects an inference backend, and provides the GitHub issue and repository. Instead of manually identifying relevant code and missing reproduction steps, she uses \repgen to restructure the report and explore the repository. The tool identifies the relevant classes and methods, constructs the execution context, and develops a reproduction plan. It then generates a candidate script and iteratively checks its syntax, API usage, and relevance to the reported failure. This process reduces the effort required to manually construct and correct a reproduction script.

\looseness=-1
Next, \repgen estimates the execution requirements, prepares a virtual environment, and installs the required dependencies. It runs the generated script and compares the observed failure with the bug report. Alice can inspect the resulting reproduction verdict and supporting evidence without manually configuring the environment or interpreting the execution output.

Finally, Alice accesses the saved \fname{reproduce.py} and executes it independently in the prepared environment. The script captures the tensor received by \textit{self.net} and compares it with the expected sanitized input. The resulting \textit{AssertionError} confirms that the network receives the original tensor containing \textit{-100}. Alice can now reuse the script to investigate the fault without repeating the repository exploration and script-generation steps.

\section{Evaluation}
\label{sec:evaluation}

We evaluated the underlying approach of \repgen through a benchmark experiment and a controlled developer study. Our technical paper reports the detailed methodology and results~\cite{shah2026imitation}. The benchmark contained 106 real-world DL bugs from 16 GitHub projects. \repgen reproduced 85 bugs, achieving a reproduction rate of 80.19\%. The best of the 24 evaluated LLM baseline configurations reproduced 60.38\%, trailing by 19.81\%. The controlled study involved 27 developers and reported a 23.35\% increase in reproduction success and a 56.8\% reduction in reproduction time with \repgen assistance. Participants also reported lower cognitive load and a higher usefulness rating (4.57 out of 5.00) compared to the baselines~\cite{shah2026imitation}. These findings indicate that \repgen can help developers complete DL fault reproduction tasks with less manual effort.

These results support the evaluated approach, while the CLI and GitHub Action provide access to its workflow and artifacts. A large-scale evaluation involving these two user interfaces remains as the scope of future work.

\section{Limitations and Threats to Validity}
\label{sec:limitations}

\repgen currently supports Python repositories and needs complete access to the required source code, data, dependencies, and execution resources. It cannot reconstruct unavailable private datasets, external services, or specialized hardware. \repgen relies on the bug report submission date to select the repository version, which may differ from the version used by the reporter. These conditions can prevent the reproduction even when the reported bug is valid.

Verification and reproduction assessment can lead to false positives sometimes. Static checks cannot establish that the reported failure occurred, and LLM assessment can misinterpret runtime output, particularly for silent bugs. Future work will make verification configurable to mitigate such challenges.

Python virtual environments separate package installations but do not isolate system drivers or all accessible files. Stronger process isolation and resource controls remain future work. Model calls and execution also consume computing resources. Exploration budgets, attempt limits, and execution decisions help control this use, but we have not measured the tool's carbon footprint, which remains future work.

\section{Related Work}
\label{sec:related-work}

Large language models (LLMs) have been used to support several software engineering tasks. For example, \textsc{LIBRO} generates regression tests from bug reports~\cite{kang2023large}, \textsc{AutoCodeRover} uses code analysis and testing for software repair~\cite{zhang2024autocoderover}, and \textsc{Bugsplainer} generates explanations for buggy code using LLMs~\cite{mahbub2023explaining}. However, these tools do not specifically address the reproduction challenges of DL systems.

For DL systems, \textsc{DeepFD} supports fault localization~\cite{cao2022deepfd}, while empirical studies highlight reproduction challenges involving training code, data, framework versions, and hardware~\cite{shah2025towards}. Silent bugs introduce further challenges because failures may occur without exceptions~\cite{tambon2024silent}. Unlike these approaches, \repgen reproduces DL bugs by combining repository-aware code generation, iterative verification, and runtime assessment. Its generated scripts and execution evidence can also support subsequent debugging activities.

\section{Conclusion}
\label{sec:conclusion}

We presented \repgen, an open-source tool for reproducing DL bugs through context-aware code generation and iterative verification. Unlike approaches that rely only on generated code or explicit failure signals, \repgen constructs a learning-enhanced context from the repository, generates reproduction scripts, and refines them using static checks and LLM feedback. It further prepares the execution environment and assesses whether the observed behaviour matches the reported failure. The underlying approach reproduced 85 of 106 real-world DL bugs and improved reproduction success and time in a controlled study with 27 developers. \repgen is available on PyPI and provides an interactive CLI for individual investigations and a GitHub Action for automated issue processing. Users can select local or hosted LLMs, configure exploration and execution budgets, and inspect generated scripts, verification findings, and reproduction evidence. These features make the tool accessible for practical DL bug investigation and support subsequent debugging activities.

\bibliographystyle{IEEEtran}
\bibliography{ref}

\end{document}